\documentclass[letterpaper,english,reprint,aps,notitlepage,superscriptaddress,twocolumn,pra,longbibliography]{revtex4-2}

\usepackage[T1]{fontenc}
\usepackage[utf8]{inputenc}
\usepackage{amsmath}
\usepackage{amssymb}
\usepackage[dvipsnames]{xcolor}
\usepackage{nicefrac}

\colorlet{mylinkcolor}{RoyalPurple}
\colorlet{mycitecolor}{RoyalPurple}
\colorlet{myurlcolor}{RoyalPurple}

\usepackage{hyperref}
\hypersetup{
	linkcolor  = mylinkcolor,
	citecolor  = mycitecolor,
	urlcolor   = myurlcolor,
	colorlinks = true,
	breaklinks = true
}

\makeatletter

\pdfpageheight\paperheight
\pdfpagewidth\paperwidth

\newcommand{\Var}{\mathrm{Var}}
\def\subfig#1{\textbf{(\lowercase{#1})}}
\makeatother

\usepackage{graphicx}
\usepackage{babel}
\usepackage{physics}

\begin{document}

\title{A Rydberg atom based system for benchmarking mmWave automotive radar chips}

\author{Sebastian Borówka}
\email{s.borowka@cent.uw.edu.pl}
\affiliation{Centre for Quantum Optical Technologies, Centre of New Technologies, University of Warsaw, Banacha 2c, 02-097 Warsaw, Poland}
\affiliation{Faculty of Physics, University of Warsaw, Pasteura 5, 02-093 Warsaw, Poland}
\author{Wiktor Krokosz}
\affiliation{Centre for Quantum Optical Technologies, Centre of New Technologies, University of Warsaw, Banacha 2c, 02-097 Warsaw, Poland}
\affiliation{Faculty of Physics, University of Warsaw, Pasteura 5, 02-093 Warsaw, Poland}
\author{Mateusz Mazelanik}
\email{m.mazelanik@cent.uw.edu.pl}
\affiliation{Centre for Quantum Optical Technologies, Centre of New Technologies, University of Warsaw, Banacha 2c, 02-097 Warsaw, Poland}
\author{Wojciech Wasilewski}
\affiliation{Centre for Quantum Optical Technologies, Centre of New Technologies, University of Warsaw, Banacha 2c, 02-097 Warsaw, Poland}
\affiliation{Faculty of Physics, University of Warsaw, Pasteura 5, 02-093 Warsaw, Poland}
\author{Michał Parniak}
\email{mparniak@fuw.edu.pl}
\affiliation{Centre for Quantum Optical Technologies, Centre of New Technologies, University of Warsaw, Banacha 2c, 02-097 Warsaw, Poland}
\affiliation{Faculty of Physics, University of Warsaw, Pasteura 5, 02-093 Warsaw, Poland}

\begin{abstract}
Rydberg atomic sensors and receivers have enabled sensitive and traceable measurements of RF fields at a wide range of frequencies. Here we demonstrate the detection of electric field amplitude in the extremely high frequency (EHF) band, at $131\ \mathrm{GHz}$. In our approach we propagate the EHF field in a beam, with control over its direction and polarization at the detector using photonic waveplates. This way, we take advantage of the highest detection sensitivity, registered for collinear propagation and circular polarization. To exhibit the potential for applications in this kind of Rydberg-atom based detection, we perform test measurements on the EHF field emitted from an on-chip radar, planned to be used in automotive industry as a vital sign detector. Our work elucidates practical applications of Rydberg-atom media as well as photonic metamaterial elements.
\end{abstract}

\maketitle

\section{Introduction}

Rydberg microwave electrometry was proposed as a means to measure weak RF fields with reference traceable to atomic transition dipole moments, the values of which can be derived from quantum atomic theory. The advantages of atomic measurements include direct measurement of E-field, intrinsic calibration, tunability to various bands, prospects for integration, and weak scattering, enabling stealthy measurements \cite{Yuan_2023,Zhang_2024}. The developments in detection with Rydberg atomic vapors enabled great sensitivity \cite{Simons_2019,Gordon_2019,Jing_2020,Borowka_2024} and expanded the measurement schemes to imaging \cite{Fan_2014,Holloway_2014} and photon-counting \cite{Bor_wka_2023}. A lot of research was devoted to analyses of atomic receivers \cite{Deb_2018,Meyer_2018,Song_2019,Holloway_2019,Jiao_2019,Simons_2019_2,Anderson_2021,Li_2022,Bor_wka_2022} and various demonstrations presented solutions adjacent to real-world applications, such as transmission of recorded sound through atomic media \cite{Holloway_2019_2}, spectral analysis \cite{Meyer_2021}, multifrequency recognition \cite{Liu_2022}, sensing at distance \cite{Otto_2023} and satellite radio reception \cite{Elgee_2023}.

The realisations presented to date involved working in the extremely high frequency (EHF) band, entering the regime of millimeter waves \cite{Gordon_2014,Thaicharoen_2019,Meyer_2023,Bohaichuk_2023,Cloutman_2024,Allinson_2023}. In this research area, specific solutions included imaging via Autler-Townes (A-T) splitting \cite{Holloway_2014} and fluorescence \cite{Downes_2023}, proposal for transduction to optical frequencies \cite{Kiffner_2016}, later realized in cryogenic vacuum \cite{Kumar_2023} and heated vapor cell \cite{Li_2024} environments and recently an EHF band receiver \cite{Legaie_2024}. 

Here we propose a measurement scheme for the calibration and testing of an on-chip sensor operating in the EHF band ($131\ \mathrm{GHz}$). Because on-chip radars are expected to be implemented as in-cabin vital sign detectors in automotive industry \cite{Antolinos_2023,Zhang_2023}, we hope that this proof-of-concept demonstration takes Rydberg atom based mmWave detection a step further towards industrial applications, allowing measurement of electric field and frequency in a band notorious for difficulties in calibration. In particular, the absolute calibration standards in the band of interest are well-defined in waveguides, relying on calorimeters \cite{Allen_1999,Gu_2019,Celep_2022}, and for free-space applications thermal blackbody calibration sources are used \cite{Houtz_2017_1,Houtz_2017_2}. Both methods enable the calibration of intensity but due to the convoluted measurement procedures serve only as primary reference standards in calibration chains. In practice, the calibrations of on-chip radars in the EHF band rely on relative measurements or simulations \cite{Sarkas_2012,Beer_2013}. Therefore, they are not absolutely calibrated and their traceability depends on the connection to the calibration chain. Rydberg atoms provide an alternative that bypasses this procedure since dipole transitions with known dipole moments can be used as standards of electric field \cite{Sedlacek_2012}. Additionally, due to many transitions in the band of interest, several independent calibration regimes are realisable with a single atomic sensor.

Furthermore, in this work we demonstrate the Rydberg atomic detection of EHF field in a configuration, where the EHF field is colinear with optical fields. In this case, the detector can be particularly sensitive to circular polarization of the EHF field,  in contrast to the perpendicular case, where the detector is polarization-insensitive \cite{Cloutman_2024}. The colinear propagation necessitates that all fields are spatially combined, which is realised with a parabolic mirror focusing the EHF field inside a rubidium vapor cell. The optimal coupling is achieved by preparing the EHF field in a collimated beam with the help of 3D-printed high impact polystyrene (HIPS) diffractive and metamaterial elements, acting as lenses and waveplates. These enable beam-shaping and control of the EHF field polarization. With the use of metamaterial waveplates, we are able to directly demonstrate the sensitivity of the system to the polarization of the detected field.

\section{Principle and Method}

\begin{figure*}
\includegraphics[width=0.9\textwidth]{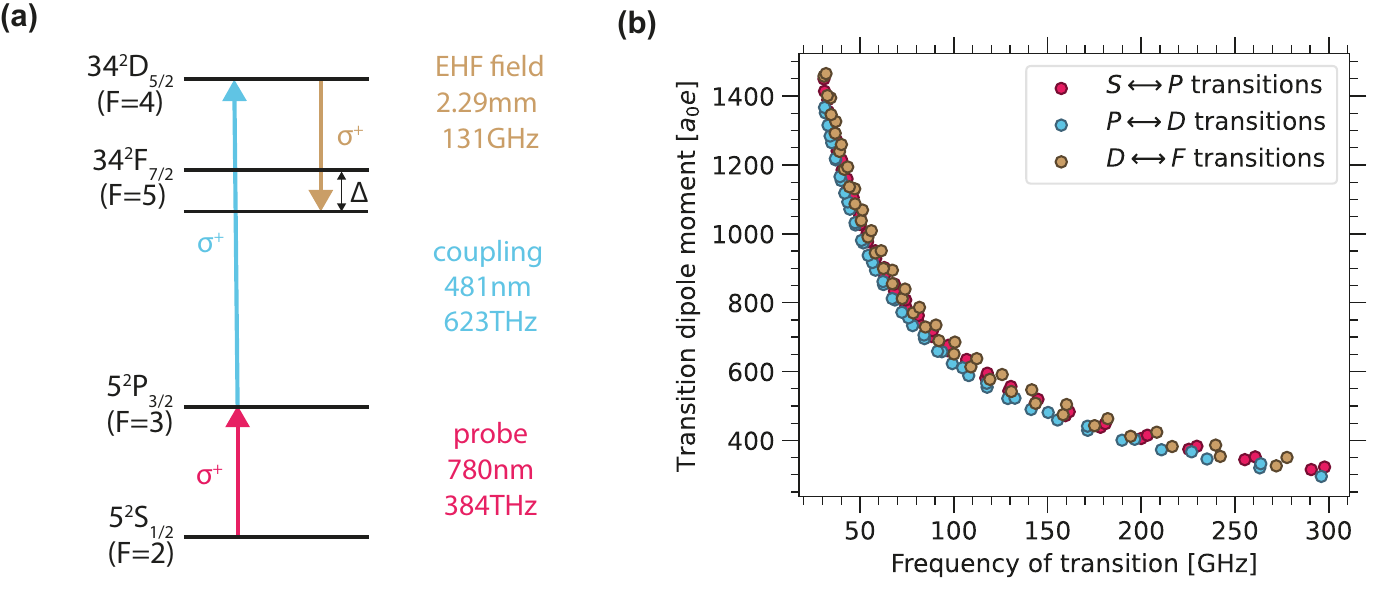}
\caption{
\subfig{a} Energy level structure utilized in the experimental setup. The two-photon (\emph{probe}-\emph{coupling}) excitation path is used to access one of the Rydberg states, enabling Rydberg transitions in the EHF regime. The probe field is scanned near the atomic resonance, while the EHF field has an additional variable detuning, indicated by the value $\Delta$ (defined here non-canonically as a detuning below the energy level, for further convenience). All of the transitions are nominally driven by matching circularly polarized fields, inducing sign-matched $\sigma$ type transitions, indicated as $\sigma^{+}$ in the subfigure. The hyperfine-split sublevels are clearly defined for $5^2\mathrm{S}_{1/2}$ and $5^2\mathrm{P}_{3/2}$ levels and inferred from having the largest transition dipole moments for $34^2\mathrm{D}_{5/2}$ and $32^2\mathrm{F}_{7/2}$ levels.
\subfig{b} The available transitions between Rydberg energy levels in rubidium, covering the whole EHF band. The plot presents the main branch (highest transition dipole moments for a given frequency) of transitions, for three families of transitions (between $\mathrm{S} \leftrightarrow \mathrm{P}$, $\mathrm{P} \leftrightarrow \mathrm{D}$ and $\mathrm{D} \leftrightarrow \mathrm{F}$ levels), including both upward and downward transitions. The plot has been generated using the results from the Alkali Rydberg Calculator library \cite{_ibali__2017}.}
\label{states}
\end{figure*}

\begin{figure*}
\includegraphics[width=0.8\textwidth]{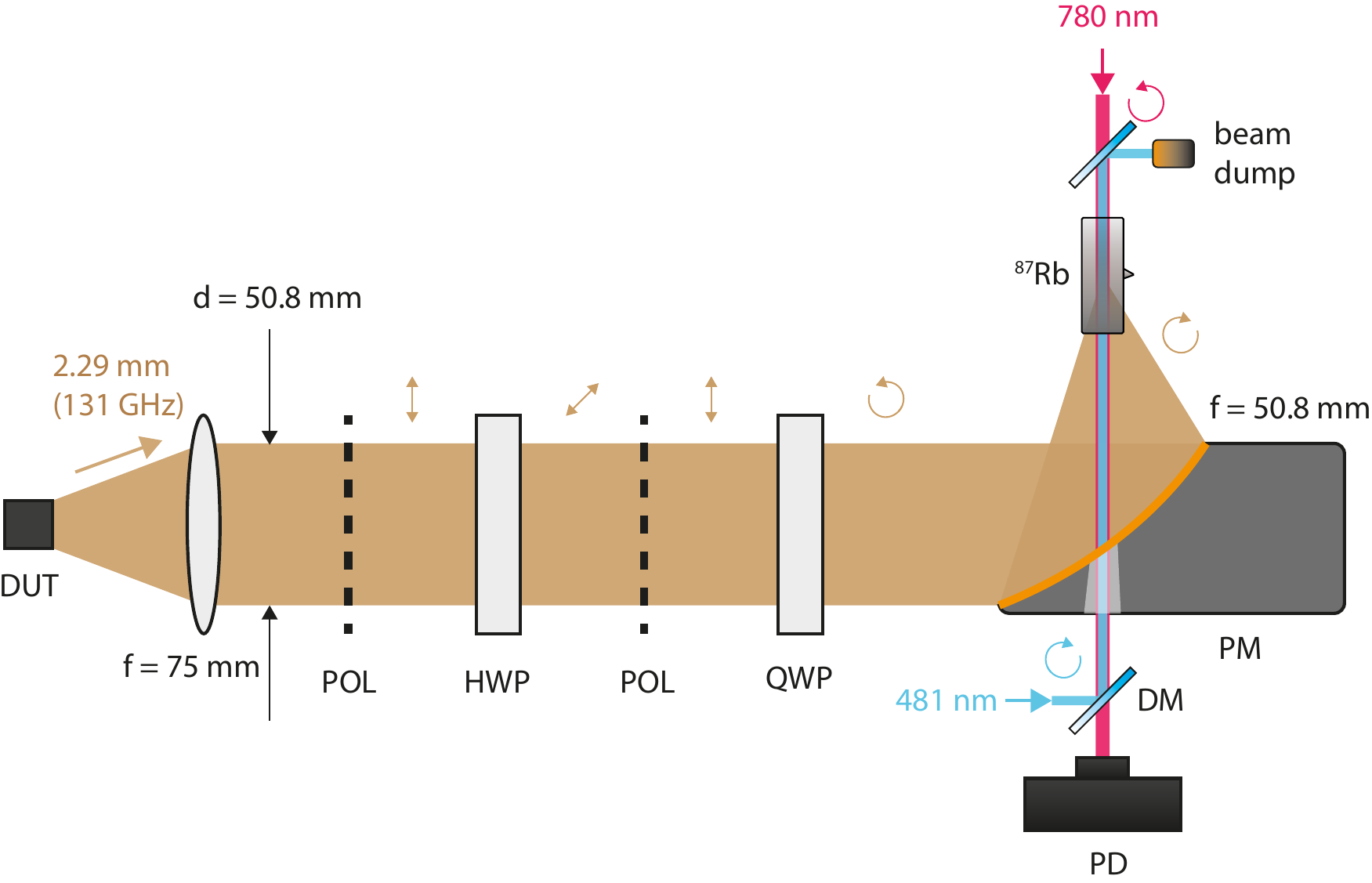}
\caption{
The setup used for the calibration of DUT (device under test). The EHF field is directionally emitted and collimated into a beam with diameter $2" = 50.8\ \mathrm{mm}$ via a lens with focal length $f = 75\ \mathrm{mm}$. The beam is passed via a setup of POL (linear polarizer), HWP (half-waveplate), POL and QWP (quarter-waveplate), allowing control of intensity and polarization of the EHF field, nominally set for circular polarization. The EHF field is focused with a PM (parabolic mirror), with reflected focal length $f = 2" = 50.8\ \mathrm{mm}$, into a quartz cylindrical ${^{87}\mathrm{Rb}}$ vapor cell ($25\ \mathrm{mm}$ optical length, $10\ \mathrm{mm}$ outer diameter, $1\ \mathrm{mm}$ thick walls, $3\ \mathrm{mm}$ thick windows). There, a probe-coupling Rydberg electrometric detection scheme is realized with counter-propagating circularly polarized optical beams combined with DM (dichroic mirrors) and focused to an interaction region corresponding to Gaussian beams with matching waists $w_0 = 250\ \mathrm{\mu m}$. The readout of probe field absorption in ${^{87}\mathrm{Rb}}$ is facilitated with a PD (photodiode).
}
\label{stp}
\end{figure*}

The detection of EHF field relies on the rubidium energy level structure visualized in the Fig.~\ref{states}\subfig{a}. The $34^2 \mathrm{D}_{5/2} \rightarrow 32^2 \mathrm{F}_{7/2}$ transition is addressed with the EHF field at $131\ \mathrm{GHz}$ ($2.29\ \mathrm{mm}$ wavelength). To access this transition, a standard probe-coupling excitation scheme is used, where absorption spectrum of scanning probe field is utilized as a detection readout. The counterpropagation of probe and coupling fields allows partial Doppler effect cancellation and enables operation in room-temperature atomic vapors. Furthermore, taking advantage of circular polarizations of fields, allows better addressing the most sensitive transitions in the degenerated hyperfine structure of the energy levels. To address various frequencies of interest, different transitions can be accessed in a similar manner. The transitions from the most sensitive branch of Rydberg transitions in rubidium are pictured in the Fig.~\ref{states}\subfig{b}, spanning throughout the whole EHF band.

To explain atom-light interaction in the depicted 4-level ladder scheme, a density matrix approach can be used, yielding particularly simple results in the form of nested Lorentz-type resonances, where weak probe field approximation is assumed \cite{Finkelstein_2023}. This approach, even expanded to the Doppler-broadened case for room-temperature atoms, shows that the A-T splitting induced by the EHF field can be directly observed in the probe field absorption as the splitting of electromagnetically induced transparency (EIT) resonance \cite{Sedlacek_2012}. In this realization, the splitting $s_{A-T}$ observed in probe field detuning can be expressed as
\begin{equation}\label{sep}
    s_{A-T} = \frac{\lambda_c}{\lambda_p} \Omega,
\end{equation}
where $\lambda_p$, $\lambda_c$ are wavelengths of probe and coupling fields and $\Omega$ is the Rabi frequency of the EHF field. This parameter is directly proportional to the amplitude of the EHF electric field $E$:
\begin{equation}\label{Om}
    \Omega = \frac{d {\cdot} E}{h},
\end{equation}
where $d$ is the transition dipole moment. In further consideration, we assume that for the chosen EHF transition $d = 542 a_0 e$ \cite{_ibali__2017}, where $a_0$ is Bohr radius and $e$ elementary charge.

Typically in Rydberg detection schemes, the EHF fields have been treated as RF fields emitted from a horn antenna. However, as the wavelengths are significantly smaller than with typical RF fields, it is possible to treat the EHF fields akin to optical or terahertz fields, that is, to propagate them in beams with finite apertures. This enables better control of the direction, which the detected field comes from, and is a first step in Rydberg atom based measurement of not only electric field amplitude, but also power. Additionally, to achieve the greatest sensitivity, control over the polarization of the EHF field is paramount. This mandates the usage of optical components that can shape, polarize and attenuate the EHF field generated by the source.

\begin{figure*}
\includegraphics[width=0.35\linewidth]{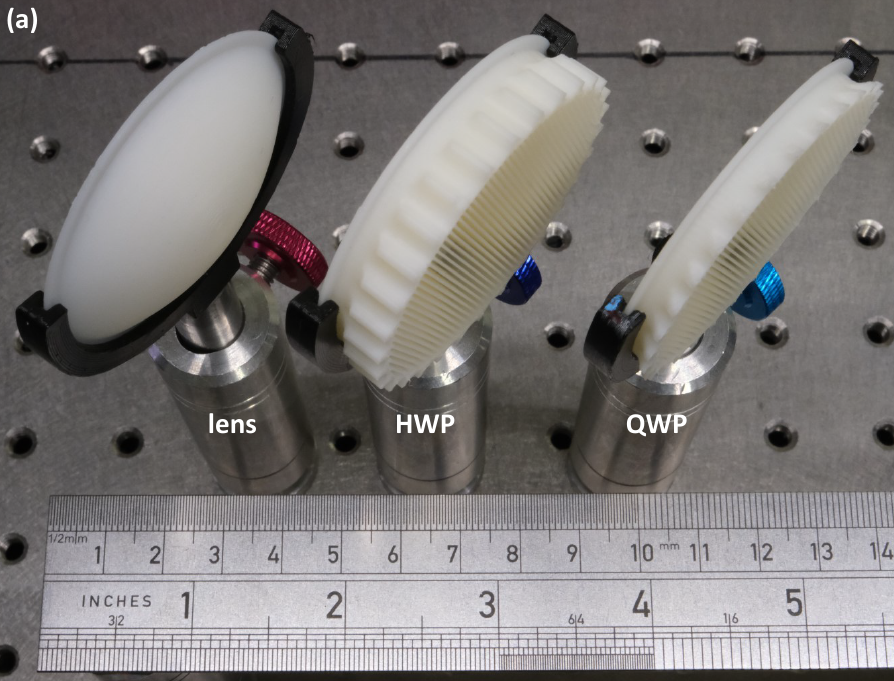}
\includegraphics[width=0.6\linewidth]{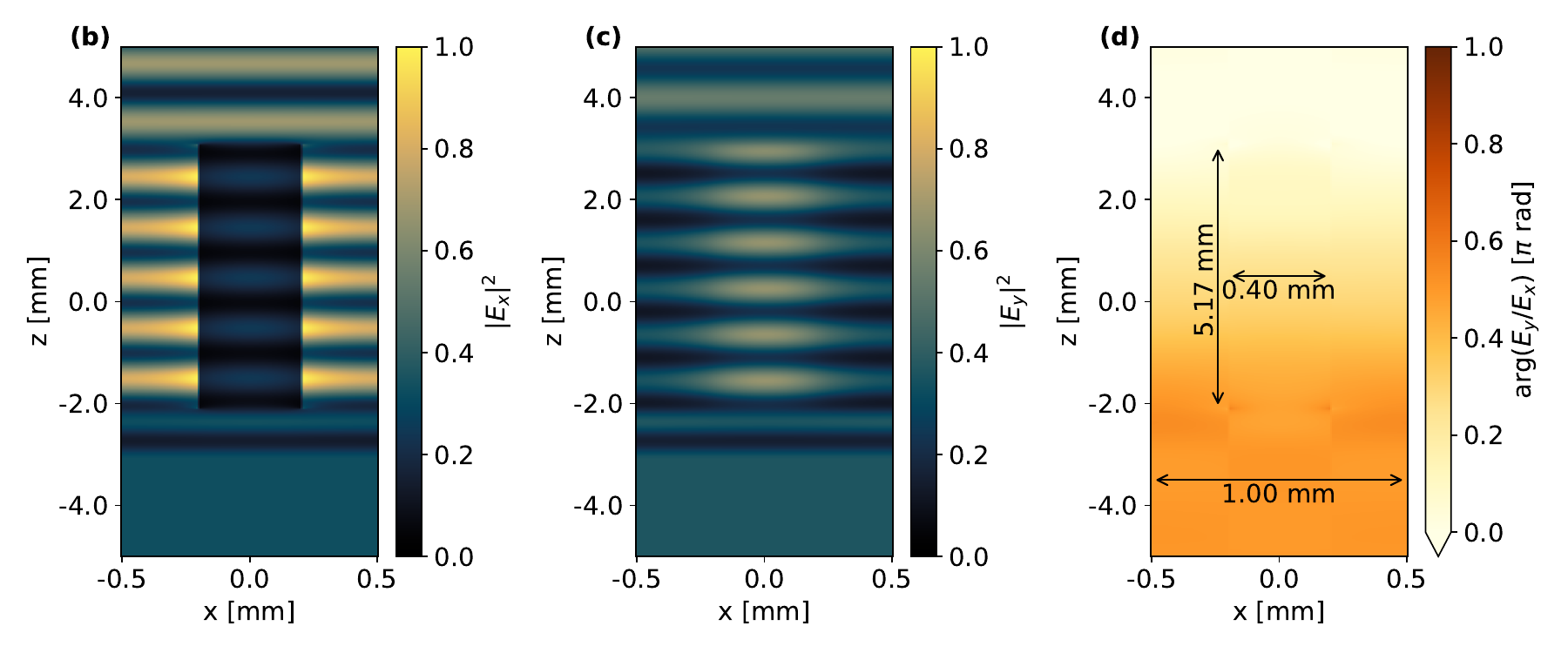}
\caption{
\subfig{a} A photograph of lens, HWP (half-waveplate) and QWP (quarter-waveplate) used in the experimental setup. The elements are 3D-printed from HIPS filament. The holder enables slide-in type mounting and rotation of waveplates, made easier by tart-like shapes of waveplates' edges. \subfig{b} Results of the 2D numerical FDTD simulation showing the $|E_x|^2$ of electric field vector in space around a single fin of the QWP, with the EHF field propagating in the $-z$ direction. \subfig{c} The corresponding results for the $|E_y|^2$. \subfig{d} The corresponding results for the phase shift between $E_y$ and $E_x$. Note the $\pi/2$ shift. The dimensions of the fin and fin pitch of the QWP are noted on the colormap. For the HWP the fin is twice as high with the other dimensions unchanged.
}
\label{fal}
\end{figure*}

\begin{figure}
\includegraphics[width=\columnwidth]{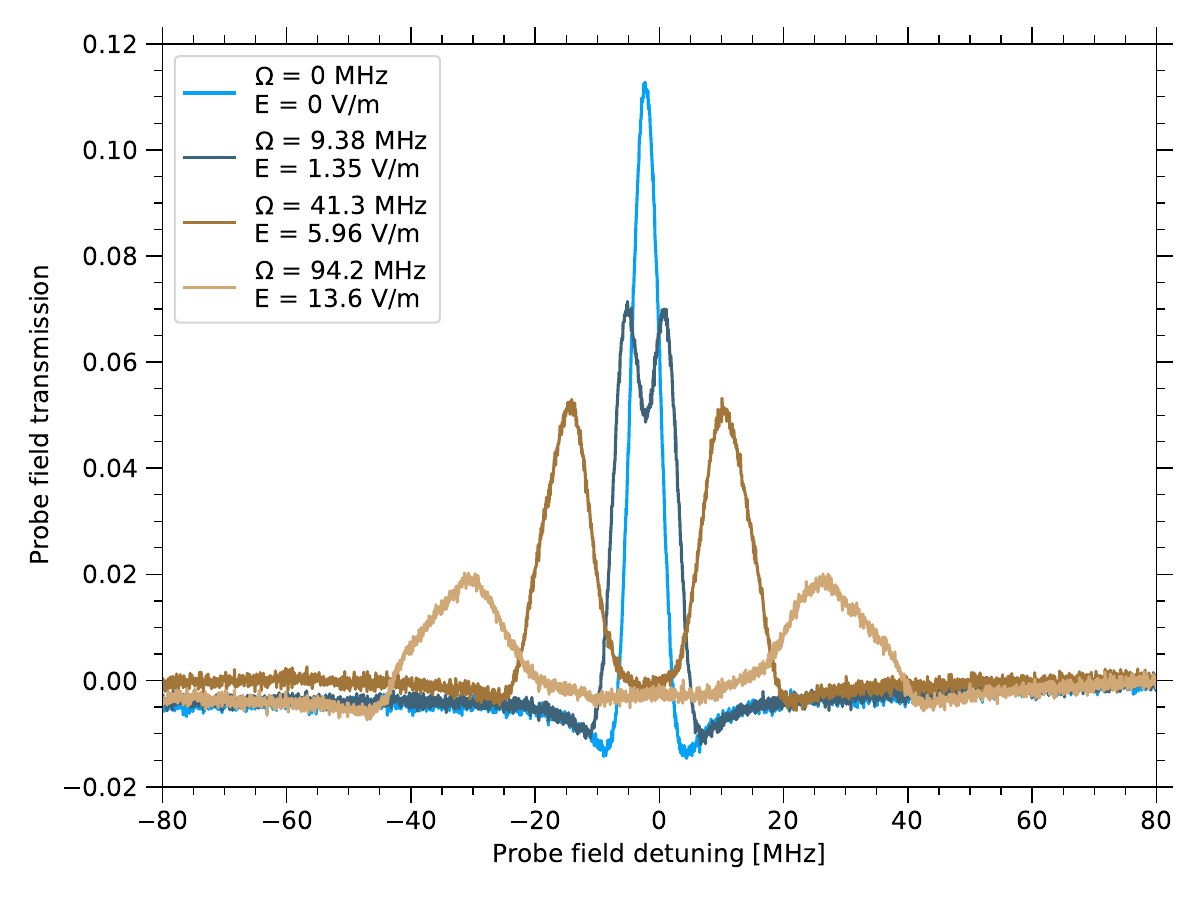}
\caption{
A-T splitting arising from the EHF field. The EHF frequency was picked to be in resonance with the Rydberg states transition, which manifests itself in the split peaks being of equal height for each signal. The visible signals present a normalized probe field transmission spectrum, where each measurement was divided by the measured background, i.e. the one-photon probe transmission spectrum. The legend presents parameters derived from the measurements, where $\Omega$ is the Rabi frequency of the EHF driven state transition and is equal to the separation between the split peaks multiplied by $\lambda_{p} / \lambda_{c}$, and $E$ is the corresponding electric field amplitude, calculated from the dipole moment relation (\ref{Om}).
}
\label{rabi}
\end{figure}

\section{Experimental setup}

The experimental setup, presented in the Fig.~\ref{stp}, relies on three fields, which are directed into a $^{87}\mathrm{Rb}$ vapor cell. The probe transmission spectrum is observed using an avalanche photodiode (Thorlabs APD430A). The coupling beam counterpropagates to offset the Doppler effect within the atomic vapors, and dichroic mirrors are utilized to combine both beams. Subsequently, the EHF field is focused inside the vapor cell with a gold-coated off-axis parabolic mirror. The mirror has a hole drilled to enable collinear introduction of optical beams. The mirror is aligned with the use of a scattering material positioned near its focal point, in place of the vapor cell. The scattered reflection of the $481\ \mathrm{nm}$ laser can be used to precisely align the focal point and angle of the mirror in relation to the interaction region. The experiment is performed in ambient Earth's magnetic field and its effect on the shifting of the magnetic sublevels is negligible.

The device under test emits the signal at a wide angle (over $20^\circ$), which necessitates the use of a collimating lens. For this purpose, a dielectric lens with a focal length of $f = 75\ \mathrm{mm}$ was 3D-printed from HIPS filament. The material has been shown to offer reasonable parameters to be used as a refractive material in the low THz regime, in particular its refractive index for EHF and THz fields is $n \approx 1.5$ \cite{Brodie2022} and its absorption coefficient $@131\mathrm{GHz}$ is around $\alpha = 0.15\ \mathrm{cm}^{-1}$, which was confirmed in our test measurements. The lens has been verified to create a collimated beam with a diameter of around $2" = 50.8\ \mathrm{mm}$. It is propagated through a linear polarizer, half-waveplate, linear polarizer and a quarter-waveplate. This part of the setup is responsible for offering continuously variable attenuation of the EHF field and inducing a circular polarization at the entry to the vapor cell. The half-waveplate as well as the following quarter-waveplate are also 3D-printed HIPS elements, photographed in the Fig.~\ref{fal}\subfig{a}, with the details concerning their shapes denoted. Their waveplate properties arise from their fin-based metamaterial structure \cite{Hernandez_Serrano_2019,Rohrbach_2021,J_ckel_2022}, designed and verified using finite difference time domain (FDTD) software \cite{tidy3D}. Figure \ref{fal}\subfig{b-d} depicts the results of the FDTD simulation, illustrating the mechanism for the birefringence. The electric field component perpendicular to the fins ($E_x$) visible in the Fig.~\ref{fal}\subfig{b} is repelled from the structure and thus experiences a lower index of refraction. At the same time, in the Fig.~\ref{fal}\subfig{c} the electric field component aligned with the fins ($E_y$) is concentrating inside the structure and experiences higher retardation, which is summarized by the induced phase difference shown in the Fig.~\ref{fal}\subfig{d}. The use of 3D-printing enables custom solutions for diffractive elements and waveplates, and is particularly low-cost, even in comparison to standard, non-custom lens and waveplates for EHF and THz fields. Here we emphasize a practical use case of these polarization elements. On the other hand, the polarizers used in the setup are PCB boards of sub-mm spaced copper paths (width $0.25\ \mathrm{mm}$ and pitch $0.6 \mathrm{mm}$).

The automotive radar chip presented as the device under test in this experiment is Indie Semiconductor TRA\_120\_045. It can be driven to emit EHF frequencies from $114$--$134\ \mathrm{GHz}$ range, though in this demonstration it is tuned narrowly around $131\ \mathrm{GHz}$ and operates in continuous wave mode. Although we are not able to straightforwardly obtain the waveform of the emitter (in which case a narrowband local oscillator field is required), with the use of an optical-bias detection, described in the Ref.~\cite{Borowka_2024}, we estimate its spectral bandwidth at ${<} 500\ \mathrm{kHz}$. In the tuning range of the device, there are overall 11 transitions pictured in the Fig.~\ref{states}\subfig{b}. The declared output power of the on-chip radar at $131\ \mathrm{GHz}$ is $3\ \mathrm{dBm}$. Its power reduction function (by $-3\ \mathrm{dB}$) and disabling the built-in amplifier (after which we still observed weak emitted field at around $-21\ \mathrm{dB}$ relative to high power mode) were used as a means to attenuate the EHF field at the detector, in addition to attenuating with a half-wave plate and polarizers. The on-chip radar is mounted on a 2-axis gimbal that allows full position and angle alignment relative to the already internally aligned setup. The alignment is performed in a way that maximizes the sensor response for low emitted power.

\begin{figure*}
    \includegraphics[width=0.75\textwidth]{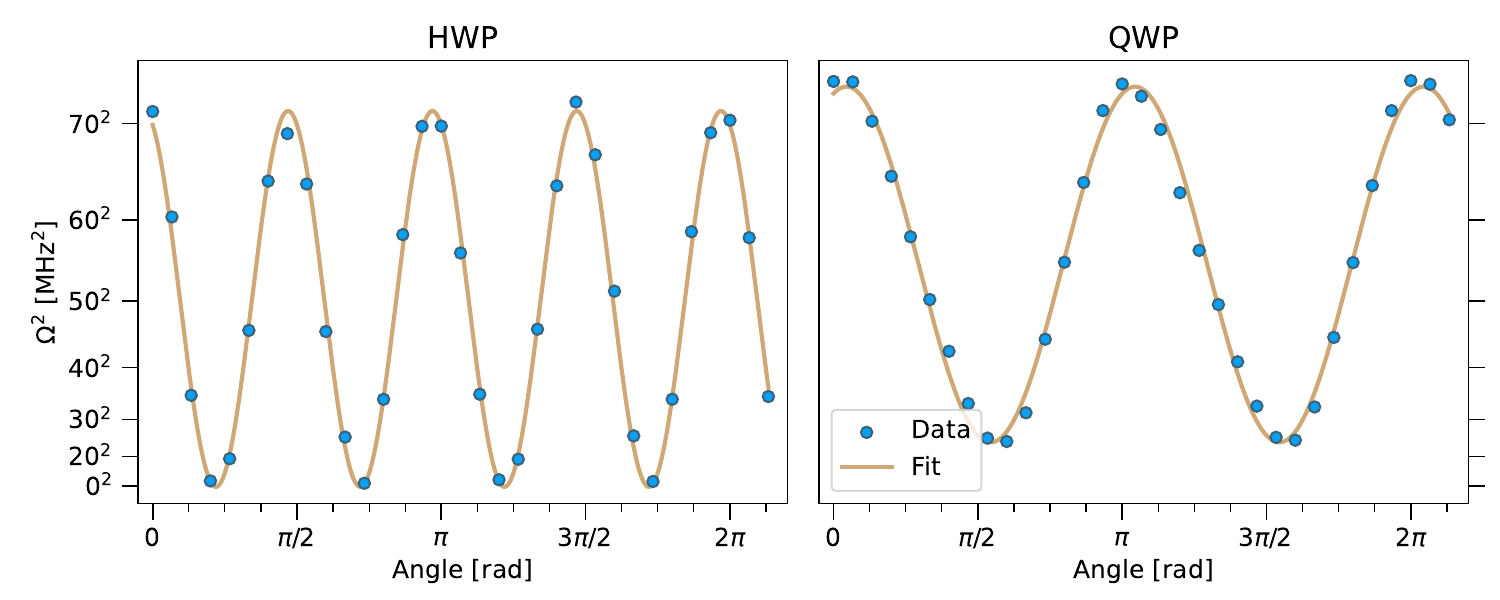}
    \caption{EHF Rabi frequencies squared, $\Omega^2$, measured via A-T splitting and corresponding to the rotation angles of the HWP and QWP in the experimental setup. We fit cosine functions to both of the results and estimate their contrasts, i.e.~the ratio between the maximum and minimum. For the HWP we obtain contrast of $420$ and for QWP -- $9.2$.
    }
    \label{waveplates}
\end{figure*}

\section{Results and Discussion}

For the demonstration of the benchmarking setup, we first measure the resonant frequency of the $34^2 \mathrm{D}_{5/2} \rightarrow 32^2 \mathrm{F}_{7/2}$ transition at $130.728\ \mathrm{GHz}$. This can be compared to the simulated prediction of $130.726\ \mathrm{GHz}$ \cite{_ibali__2017}. At this resonant frequency we perform a standard A-T splitting measurement of the EHF electric field for various levels of this field, applying the relations (\ref{sep}) and (\ref{Om}). These results are presented in the Fig.~\ref{rabi} in the domain of probe field detuning, where zero detuning is defined as two-photon EIT resonance.

For strong fields the split peaks are broadening, which we attribute to the EHF field inhomogeneity inside the vapor cell. We estimate the inhomogeneity to be $7.3\%$ in the intensity relative units (see Appendix A for the full consideration). On the other hand, for weak fields the splitting becomes unresolvable with the standard technique, therefore, other methods have to be used for an absolute calibration \cite{Jing_2020,Borowka_2024}. We estimate the weakest resolvable field to be $0.65\ \mathrm{V}/\mathrm{m}$, corresponding to the Rabi frequency $\Omega_0 = 4.5\ \mathrm{MHz}$ (see Appendix B for the full consideration). Furthermore, we estimate that the achievable sensitivity of detection in the presented setup is $9.8\ \mathrm{\mu V}/\mathrm{m}/\sqrt{\mathrm{Hz}}$ in the zero detuning point -- for weak fields but simultaneously in the regime, where the A-T splitting is still resolvable (see Appendix C for the full consideration).

Next, we use the setup to estimate the overall efficiency of the control over polarization. We rotate the HWP and QWP in the experimental setup and measure the Rabi frequency with the A-T splitting method, assuming $d = 542 a_0 e$ for the $34^2\mathrm{D}_{5/2} (\mathrm{F}{=}4,\mathrm{m}_{\mathrm{F}}{=}4) \rightarrow 32^2\mathrm{F}_{7/2} (\mathrm{F}{=}5,\mathrm{m}_{\mathrm{F}}{=}5)$ transition. The transitions to the outlier hyperfine magnetic states can be assumed due to the use of circular polarizations of the optical beams, and having the largest transition dipole moments at each transition. Thus, due to the accumulating effect of subsequent dipole moments, only one transition is addressed efficiently, even though the magnetic levels are degenerated in the ambient magnetic field. This is evidenced by the observation of a single A-T splitting, broadening spectrally mainly due to field inhomogeneity. The results are presented in the Fig.~\ref{waveplates}. Note that the results are presented as $\Omega^2$, effectively in the domain of field intensities. We fit cosine functions to the results to estimate the contrast and visibility. For the HWP we arrive with the contrast (ratio between the maximum and minimum) of $420$ ($26\ \mathrm{dB}$), that translates to the interferometric visibility of $0.995$. For the QWP we get the contrast of $9.2$. The minimal achievable splitting remains substantial, which is the result of the A-T splitting still being induced, although via a different transition, induced by the reverse polarization, with different dipole moment. Although various transitions may be considered, and the definite answer requires the full consideration of the state evolution model, the most obvious candidate is the $34^2\mathrm{D}_{5/2} (\mathrm{F}{=}4,\mathrm{m}_{\mathrm{F}}{=}4) \rightarrow 32^2\mathrm{F}_{7/2} (\mathrm{m}_{\mathrm{J}}{=}3/2)$ transition, having a dipole moment of $d' = 118 a_0 e$ \cite{_ibali__2017}. With that assumption, the contrast for a perfect QWP should yield $21$. In that case, in relation to the maximal contrast, we achieve the visibility of $0.88$.

\begin{figure}
\includegraphics[width=\columnwidth]{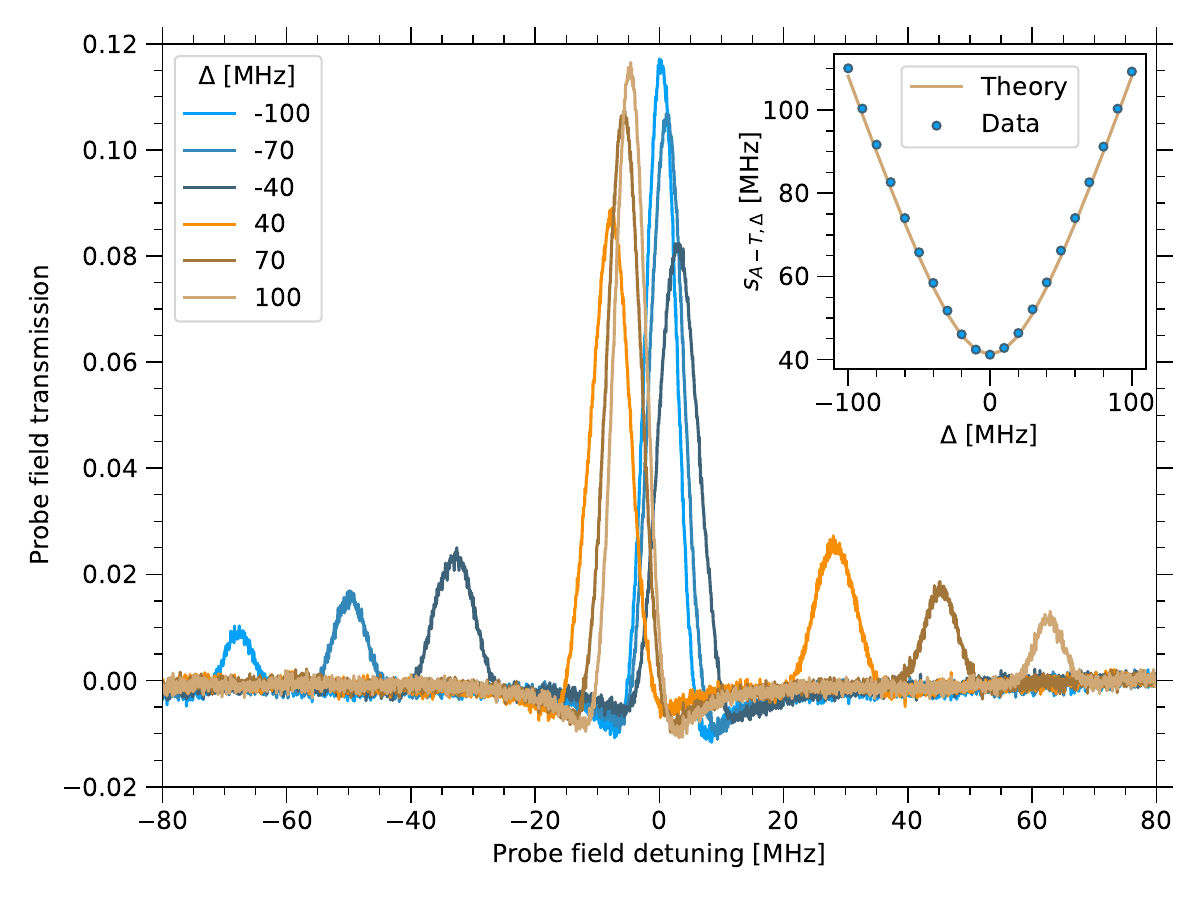}
\caption{
A-T splitting arising from various detuned EHF fields. The Rabi frequency is constant throughout the series and measured at $\Delta = 0$ as $\Omega = 41\ \mathrm{MHz}$.Each line corresponds to a different EHF field detuning $\Delta$, which results in increasing splitting between the peaks, as well as increasing difference in the height of the peaks. All signals were normalized in the same manner as was the case with the results in the Fig.~\ref{rabi}. The inset presents results of peak separation $s_{A-T,\Delta}$ measurements in relation to $\Delta$. The measurements are compared to the theoretical estimation based on the Eq.~(\ref{detsep}), yielding an average deviation of $1.1\ \mathrm{MHz}$.
}
\label{det}
\end{figure}

Consequently, we demonstrate that using this setup, an estimation of EHF field's frequency, in particular the $\Delta$ detuning, is possible as well. For the set EHF electric field, we perform measurements for different $\Delta$ detunings, and present the results for chosen detunings in the Fig.~\ref{det}. In this case, a crude estimation is provided by the relation describing the separation between the peaks
\begin{equation}\label{detsep}
    s_{A-T,\Delta} = \frac{\lambda_c}{\lambda_p} \sqrt{\Omega^2 + \Delta^2},
\end{equation}
where the parameters are defined as in the Eq.~(\ref{sep}), with $\Delta$ corresponding to the detuning from the Fig.~\ref{states}\subfig{a}. The comparison between the presented relation and obtained results is presented in the inset to the Fig.~\ref{det}. In this consideration $\Omega$ is considered set and constant, measured at $\Delta = 0$ detuning as $\Omega = 41\ \mathrm{MHz}$. When the set $\Delta$ are taken as ground truth, the estimation results from the $\pm 100\ \mathrm{MHz}$ range yield an average deviation of $1.1\ \mathrm{MHz}$. This method is not suitable for precise estimation of frequency, that instead should be obtained from wavemixing with a known local oscillator \cite{Gordon_2019}, however, it gives a reasonable estimate. The limits of this method point to the available bandwidth of the detection, which in the case of FWHM consideration is $93\ \mathrm{MHz}$ at $\Omega = 41\ \mathrm{MHz}$ (see Appendix D for the full consideration).

\section{Summary and Perspectives}

This proof-of-concept demonstration underlines the control of EHF field direction and polarization, and presents the detection scheme in a collinear configuration. The practicality of this approach is further emphasised by the demonstration of a calibration procedure for an automotive radar chip, bringing this technique a step closer towards real-world applications. Our work demonstrates a unique combination of the usage of Rydberg media, photonic metamaterial elements in a practical application. There are a few aspects that escaped the scope of this work, though. In particular, the absolute estimation of measured EHF power is missing. This is, however, a persisting problem, coming down to the estimation of the relation between the atomic interaction region and region, where the EHF field is focused.

Furthermore, the spectral region for detection using the splitting of EIT is limited by the Voigt profile of the probe absorption line at room temperature. While for lower frequencies this suffices to completely cover the E-M spectrum in a semi-continuous fashion, in the EHF band the gaps in the spectrum start to appear. These gaps can be addressed by accessing transitions from less efficient branches (smaller transition dipole moments), using different atomic media (sodium, potassium, cesium), or abandoning absolute measurements for certain frequencies, while relying on relative measurements in conventional setups.

We anticipate further development will be focused around adding resonant structures to the detection setup. This may enhance the detection sensitivity and provide an additional degree of freedom for frequency tuning. This is particularly feasible, as on the one hand for the EHF wavelengths the sizes of the resonant structures, e.g.~cavities, can be manageable around a typical optical setup, on the other hand, the manufacturing precision required is low enough that the elements allow for integration with optics, e.g.~with through-holes for optical beams.

In principle, this detection setup can be repurposed for other modes of operation, such as single-photon counting, enabled by EHF-to-optical conversion. We expect that due to the predicted directionality of the converting receiver, it can be used efficiently for quantum temperature detection in the EHF band, as the thermal radiation in the super high frequency (SHF) band has already been directly observed in a similar system  \cite{Bor_wka_2023}.

\section*{Data availability}
Data underlying the results presented in this paper are available in the Ref.~\cite{dataverse}.

\section*{Code availability}
The codes used for the numerical simulation are available from M.M.~upon request.

\begin{acknowledgments}
We thank K.~Banaszek for the generous support. This research was funded in whole or in part by National Science Centre, Poland grant no. 2021/43/D/ST2/03114.
\end{acknowledgments}

\section*{Appendix A: estimation of the field inhomogeneity}

The inhomogeneity of the EHF field inside the vapor cell can be assumed as purely scalar (field intensity) and only in the direction of the beam propagation (the inhomogeneity in the perpendicular directions is negligible due to the interaction area being subwavelength). This inhomogeneity can be estimated by measuring the dependence of the width of the A-T split peaks on the Rabi frequency of the EHF field. There we expect a linear relation
\begin{equation}\label{fit}
    \sigma \frac{\lambda_c}{\lambda_p} = \sigma_0 + \alpha \Omega,
\end{equation}
where $\sigma$ is the measured width of the A-T split peaks (in the units of probe detuning), $\sigma_0$ and $\alpha$ are parameters of the fit, and the $\frac{\lambda_c}{\lambda_p}$ factor appears due to the Doppler effect.

\begin{figure}
    \includegraphics[width=\columnwidth]{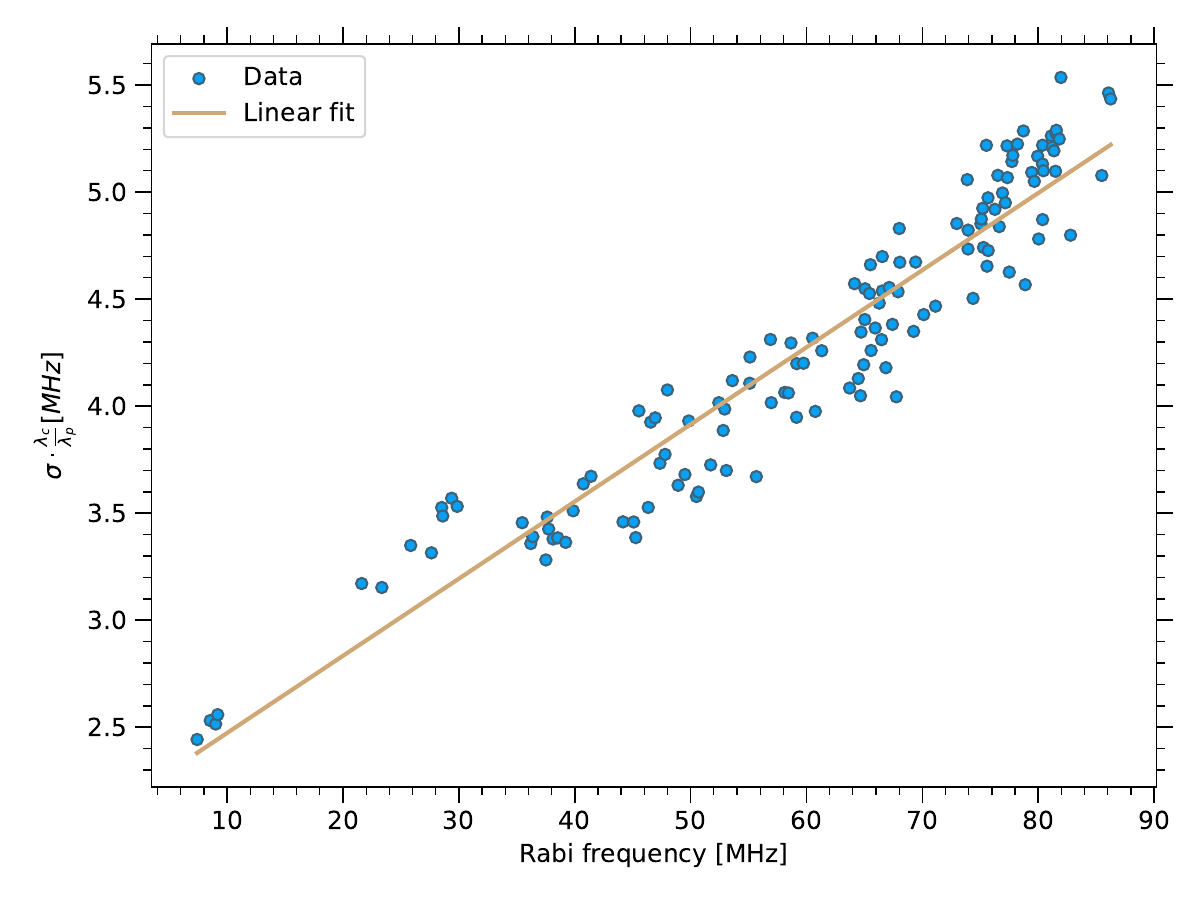}
    \caption{The analysis of the EHF field inhomogeneity based on the relation between the width of the A-T split peaks and the Rabi frequency $\Omega$ of the EHF field. The obtained parameters of the linear fit described in (\ref{fit}) are $\alpha = 0.036$, $\sigma_0 = 2.1\ \mathrm{MHz}$.}
    \label{inhm}
\end{figure}

We take data from testing of different manufactured waveplates to analyse this relation by performing Gaussian fits to the A-T split peaks and extracting the separation (translating to $\Omega$) and standard deviation (translating to the width of the peaks being the result of the inhomogeneity). The results are presented in the Fig.~\ref{inhm}. We obtain the parameters $\alpha = 0.036$, $\sigma_0 = 2.1\ \mathrm{MHz}$. The inhomogeneity (in the units of intensity) can be then calculated as $\alpha(2+\alpha) = 0.073 = 7.3\%$.

Note that the potential effects of other magnetic levels taking part in the A-T splitting experience the same linear scaling. However, as we use sign-matched circular polarizations of fields and address mostly the outlier transitions between maximal/minimal magnetic number states, these effects can be neglected. Furthermore, because the states of lower magnetic numbers experience weaker A-T splitting, their influence would overall result in a widening of the A-T split peaks in the direction of the center (zero probe field detuning). This is evidently not the case, as pictured in the Fig.~\ref{rabi} for stronger EHF fields.

\section*{Appendix B: estimation of the weakest resolvable field}

Incidentally, the fit described in the (\ref{fit}) can serve as the basis for the estimation of weakest resolvable field in the presented system. Let us consider the Rayleigh $2 \sigma$ criterion for distinguishing two separated Gaussian peaks. In the case of the A-T splitting this can be expressed as
\begin{equation}
    s_{A-T} = 2 \sigma \Longrightarrow \Omega_0 = 2 (\sigma_0 + \alpha \Omega_0),
\end{equation}
where we can extract the $\Omega_0 = 4.5\ \mathrm{MHz}$ from. This corresponds to the weakest resolvable field of $0.65\ \mathrm{V}/\mathrm{m}$ for the absolute calibration (A-T splitting technique).

\section*{Appendix C: estimation of the sensitivity}

To estimate the achievable sensitivity in the presented system, a few parameters need to be considered. Let us denote by $T$ the normalized transmission of the probe field as presented in the Fig.~\ref{rabi}. The sensitivity to electric field change $\zeta$ in a single-point measurement lasting for time $t$ is equal to
\begin{equation}\label{zeta}
    \zeta = \sqrt{t \times \Var(T)} \frac{1}{\frac{\partial T}{\partial E}} = \sqrt{t \times \Var(T)} \frac{h}{\frac{\partial T}{\partial \Omega} \cdot d},
\end{equation}
where $\Var(T)$ is the variance of the measured transmission and $\frac{\partial T}{\partial E}$ is the derivative of the transmission over the electric field. We further apply the (\ref{Om}) relation to obtain the derivative over $\Omega$.

From the measurements of noise we obtain $\sqrt{t \times \Var(T)} = 1.5{\cdot}10^{-6} /\sqrt{\mathrm{Hz}}$, which we identify experimentally as the shot noise of transmitted probe field. The $\frac{\partial T}{\partial \Omega}$ is significantly dependent on the value of $\Omega$, which is one of the reasons why the best sensitivity is achieved with the use of a local oscillator field biasing the detector to the optimal working point \cite{Jing_2020}. We can measure the derivative experimentally only in the region where the A-T splitting is resolvable. This is not, however, the most sensitive working point.

Let us first consider this problem on a theoretical model with the assumption of a weak probe field. The probe field normalized transmission in this case can generally be expressed as \cite{Finkelstein_2023}
\begin{equation}
    T = 1 - \frac{2 \Delta_1 + i \Gamma_1}{2 \Delta_1 + i \Gamma_1 - \frac{\Omega_c^2}{2 \Delta_2 + i \Gamma_2 - \frac{\Omega^2}{2 \Delta_3 + i \Gamma_3}}},
\end{equation}
where $\Delta_x$ and $\Gamma_x$ are the detunings and decay rates (including other sources of decay, such as transit time-broadening), and $\Omega_c$ is the Rabi frequency of coupling field and $\Omega$ is defined as in the main text. In the limit of a weak electric field $E$, and thus small A-T splittings we can assume $\Delta_1=0$, as this is the most sensitive point in the transmission spectrum $T$. Let us further simplify this model by assuming a completely resonant case, that is $\Delta_x = 0$, and neglecting the Doppler effect, thus considering only the velocity class, where the velocity of atoms $v = 0$ (although taking into account the transit-time broadening in $\Gamma_x$). Then
\begin{equation}
    T = 1 - \frac{\Gamma_1 (\Gamma_2 \Gamma_3 + \Omega^2)}{\Gamma_1 (\Gamma_2 \Gamma_3 + \Omega^2) + \Gamma_3 \Omega_c^2}.
\end{equation}
Following that
\begin{equation}\label{der}
    \frac{\partial T}{\partial \Omega} = \frac{2 \Gamma_1 \Gamma_3 \Omega_c^2 \Omega}{(\Gamma_3 \Omega_c^2 + \Gamma_1(\Gamma_2 \Gamma_3 +\Omega^2))^2}.
\end{equation}

After substituting the assumed and measured experimental values, $\Omega_c = 4\ \mathrm{MHz}$, $\Gamma_1 = 7.3\ \mathrm{MHz}$, $\Gamma_2 = \Gamma_3 = 1.2\ \mathrm{MHz}$ (dominated by transit-time broadening), we obtain the relationship, from which we extract the optimal working point at $\Omega = 1.2\ \mathrm{MHz}$. By inserting the results to the (\ref{zeta}), we obtain the achievable sensitivity of $\zeta = 1\ \mathrm{\mu V}/\mathrm{m}/\sqrt{\mathrm{Hz}}$. However, as the optimal working point falls outside the resolvable regime, we are able to demonstrate the sensitivity only for higher Rabi frequencies. There, for resolvable but still relatively low ($\Omega < 10\ \mathrm{MHz}$) fields we obtain an average sensitivity of $\zeta = 9.8\ \mathrm{\mu V}/\mathrm{m}/\sqrt{\mathrm{Hz}}$. Note that these values are realistically obtainable only in the superheterodyne setup with an additional LO field, as the model used to derive them does not account for probe power noise (especially long-term fluctuations, as the $\Var(T)$ is obtained during a single scan).

\begin{figure}
    \includegraphics[width=\columnwidth]{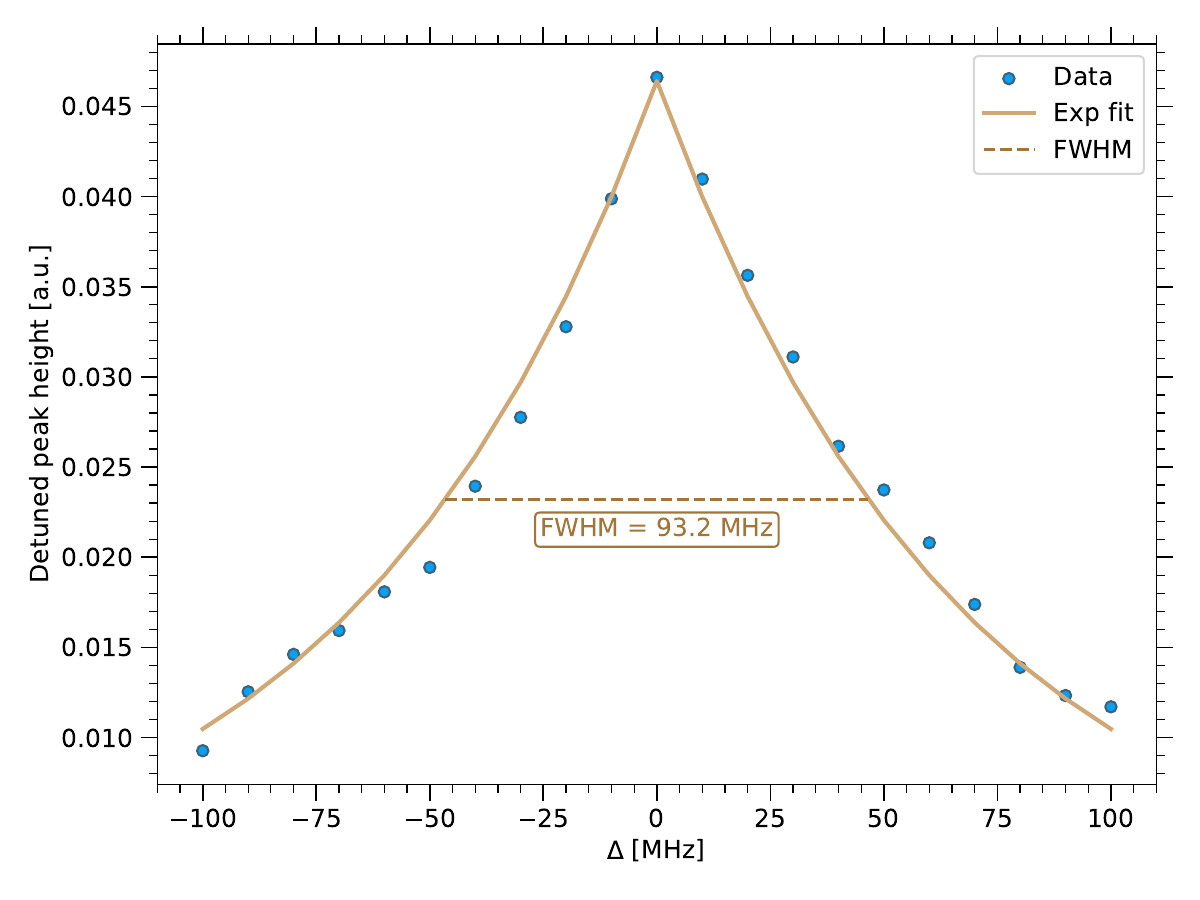}
    \caption{The analysis of the bandwidth in the presented mode of detection (A-T splitting). The height of the detuned peak is dependent on the detuning $\Delta$ of the EHF field. The exponential function is fit to this relation and the bandwidth parameter is extracted as $93\ \mathrm{MHz}$ of FWHM bandwidth.}
    \label{detpk}
\end{figure}
\section*{Appendix D: estimation of the detection bandwidth}

The further analysis of the data presented in the Fig.~~\ref{det} leads to obtaining a bandwidth available in the presented mode of detection, for a given Rabi frequency of $\Omega = 41\ \mathrm{MHz}$ in this case. In particular, analysis of the detuned peak height dependence on the detuning $\Delta$ leads to the estimation of how well the A-T splitting can be interpreted for an off-resonant EHF field. The results of this analysis are presented in the Fig.~\ref{detpk}, where we perform a fit of exponential function to extract the relevant parameters. The FHWM bandwidth in this case is $93\ \mathrm{MHz}$. From the fit we can additionally estimate the detunings, at which the detuned peak approaches the noise level (for the optimally chosen frequency binning) -- in this case the $\mathrm{SNR} = 1$ bandwidth is $330\ \mathrm{MHz}$.

\bibliography{refs}

\end{document}